\documentclass{article}

\usepackage{arxiv}

\usepackage[utf8]{inputenc} % allow utf-8 input
\usepackage[T1]{fontenc}    % use 8-bit T1 fonts
\usepackage{hyperref}       % hyperlinks
\usepackage{url}            % simple URL typesetting
\usepackage{booktabs}       % professional-quality tables
\usepackage{amsfonts}       % blackboard math symbols
\usepackage{nicefrac}       % compact symbols for 1/2, etc.
\usepackage{microtype}      % microtypography
\usepackage{lipsum}
\usepackage{graphicx}
\graphicspath{ {./images/} }
\usepackage{graphicx}%
\usepackage{multirow}%
\usepackage{amsmath,amssymb,amsfonts}%
\usepackage{amsthm}%
\usepackage{mathrsfs}%
\usepackage[title]{appendix}%
\usepackage{xcolor}%
\usepackage{textcomp}%
\usepackage{manyfoot}%
\usepackage{booktabs}%
\usepackage{algorithm}%
\usepackage{algorithmicx}%
\usepackage{algpseudocode}%
\usepackage{listings}%
\usepackage[bottom]{footmisc}
\usepackage{authblk}

\title{DomURLs\_BERT: Pre-trained BERT-based Model for Malicious Domains and URLs Detection and Classification}

\author[1]{\textbf{Abdelkader El Mahdaouy}}
\author[2]{\textbf{Salima Lamsiyah}}
\author[1]{\textbf{Meryem Janati Idrissi}}
\author[3]{\textbf{Hamza Alami}}
\author[4,5]{\textbf{Zakaria Yartaoui}}
\author[1]{\\\textbf{Ismail Berrada}}

\affil[1]{College of Computing, Mohammed VI Polytechnic University, Ben Guerir, Morocco}
\affil[2]{Department of Computer Science, Faculty of Science, Technology and Medicine,University of Luxembourg, Luxembourg}
\affil[3]{LISAC Laboratory, Faculty of Sciences Dhar El Mehraz, USMBA, Fez, Morocco}
\affil[4]{Vanguard Center, Mohammed VI Polytechnic University, Ben Guerir, Morocco}
\affil[5]{National Moroccan Computer Emergency and Response Team (maCert), Morocco\\ 

\texttt{firstname.lastname@\{um6p.ma$^{1,4}$|uni.lu$^2$|usmba.ac.ma$^3$\}}}

\begin{document}
\maketitle
\begin{abstract}
Detecting and classifying suspicious or malicious domain names and URLs is fundamental task in cybersecurity. To leverage such indicators of compromise, cybersecurity vendors and practitioners often maintain and update blacklists of known malicious domains and URLs. However, blacklists frequently fail to identify emerging and obfuscated threats. Over the past few decades, there has been significant interest in developing machine learning models that automatically detect malicious domains and URLs, addressing the limitations of blacklists maintenance and updates. In this paper, we introduce DomURLs\_BERT, a pre-trained BERT-based encoder adapted for detecting and classifying suspicious/malicious domains and URLs. DomURLs\_BERT is pre-trained using the Masked Language Modeling (MLM) objective on a large multilingual corpus of URLs, domain names, and Domain Generation Algorithms (DGA) dataset.  In order to assess the performance of DomURLs\_BERT, we have conducted experiments on several binary and multi-class classification tasks involving domain names and URLs, covering phishing, malware, DGA, and DNS tunneling. The evaluations results show that the proposed encoder outperforms state-of-the-art character-based deep learning models and cybersecurity-focused BERT models across multiple tasks and datasets.  The pre-training dataset\footnote{\url{https://hf.co/datasets/amahdaouy/Web_DomURLs}}, the pre-trained DomURLs\_BERT\footnote{\url{https://hf.co/amahdaouy/DomURLs\_BERT}} encoder, and the experiments source code\footnote{\url{https://github.com/AbdelkaderMH/DomURLs_BERT}} are publicly available. 
\end{abstract}

% keywords can be removed

\section{Introduction}
Domain names and Uniform Resource Locators (URLs) are fundamental components in navigating and identifying resources on the Internet. Nevertheless, they are frequently exploited for various malicious activities in cyberspace, such as phishing campaigns, malware distribution, spam dissemination, and Command and Control (C\&C) server operations, among others \cite{yadav2010detecting,sahoo2017malicious,kang2021,aljabri2022detecting,yu2024efficient}.Thus, detecting and flagging malicious domains and URLs is crucial for network security. Traditionally, cybersecurity vendors and practitioners rely on blacklists and heuristic methods to identify malicious domain names and URLs \cite{le2018urlnet,catal2022applications,shi2018malicious,cucchiarelli2021algorithmically}. While blacklists are essential for blocking known threats, they are reactive by nature, posing challenges in maintenance and being vulnerable to evasion techniques. On the other hand, heuristic methods, which use patterns and behavioral analysis to identify potential threats, offer a more proactive approach to detection \cite{catal2022applications,da2020heuristic,hamroun2024review}. However, they are prone to false positives and require continuous updates to remain effective against evolving obfuscation tactics \cite{aljabri2022detecting,yu2024efficient,da2020heuristic,tuan2023utldga22,bozkir2023grambeddings}.

To overcome the limitations of blacklisting and heuristic-based methods, a growing body of research has focused on developing Machine Learning (ML) techniques for detecting malicious URLs and domain names \cite{sahoo2017malicious,shi2018malicious,tuan2023utldga22}. The goal of these approaches is to automatically train models that can distinguish between legitimate and malicious threats based on data. Traditional ML-based techniques rely heavily on hand-engineered features, where the learning process involves identifying patterns in the data to guide the model's decision-making. Consequently, numerous studies have proposed various feature sets for ML-based classification of malicious domains and URLs \cite{sahoo2017malicious,aljabri2022detecting,cucchiarelli2021algorithmically,hamroun2024review}. Although ML-based methods have demonstrated promising results across different domain name and URL classification tasks, the manual feature engineering process is both costly and time-consuming \cite{le2018urlnet,catal2022applications}.

Recently, a considerable amount of literature has been published on the use of Deep Learning (DL) for detecting and classifying malicious domain names and URLs \cite{yu2024efficient,le2018urlnet,catal2022applications,bozkir2023grambeddings}. These studies leverage the representation learning capabilities of deep neural networks, which can automatically learn hierarchical features at different levels of abstraction from raw input data \cite{lecun2015deep}. As a result, various neural network architectures have been explored. Typically, these architectures either use hand-engineered features or learn representations of characters, n-grams, and sub-words for classifying malicious domain names and URLs \cite{le2018urlnet,bozkir2023grambeddings,vazhayil2018comparative,afzal2021urldeepdetect,LiewL23}.

The introduction of the transformer architecture \cite{VaswaniSPUJGKP17} has resulted in significant breakthroughs and advancements in Artificial Intelligence. Beyond natural language processing, transformers have been employed in various fields such as computer vision, data science, robotics, and cybersecurity \cite{SecBERT,aghaei2023securebert,BayerKSR24,xu2024large}. Particularly, self-supervised pre-training of stacked transformer blocks—whether in encoder, decoder, or encoder-decoder configurations—has demonstrated state-of-the-art performance when fine-tuned for downstream tasks \cite{zhao2023survey,minaee2024llm}. In line with the pretrain-finetune paradigm, researchers have proposed fine-tuning or adapting pre-trained Bidirectional Encoder Representations from Transformers (BERT) \cite{Devlin2019BERTPO} for cybersecurity tasks \cite{yu2024efficient,BayerKSR24,xu2024large,chang2021research,otieno2023detecting,su2023bert,li2024URLBERT,liu2023malicious,ICASSP2023,tian2024dom}. Following the domain-adaptive pre-training approach, several BERT-based encoders have been pre-trained using the Masked Language Modeling (MLM) objective on domain-specific corpora for the classification of malicious and phishing URLs \cite{xu2024large,li2024URLBERT,URLTran9653028,ICASSP2023}. However, these models have not been explicitly pre-trained on both domain names and URLs, and much of the existing research has focused primarily on phishing URLs.

In this paper, we introduce {DomURLs\_BERT}, a BERT-based encoder pre-trained on a large-scale corpus using the MLM objective. The pre-training corpus includes multilingual URLs, domain names, and Domain Generation Algorithms (DGA) datasets. Additionally, we propose a lightweight preprocessing method for the input data and train our model's tokenizer from scratch using SentencePiece tokenization. To evaluate the performance of {DomURLs\_BERT} in detecting malicious URLs and domain names, we conducted a comprehensive evaluation on a diverse set of datasets covering DGA, DNS tunneling techniques, malware classification, and phishing/malicious URL classification. The overall results show that our model outperforms six character-based deep learning models and four BERT-based models on multiple classification tasks. To summarize, the main contributions of this paper are as follows:

\begin{itemize}
    \item  We introduce {DomURLs\_BERT}, a specialized BERT-based encoder pre-trained on a large-scale multilingual corpus of URLs, domain names, and DGA datasets.
    \item We propose a light preprocessing tailored to the characteristics of URLs and domain names, and train domain-specific tokenizer.
    \item We evaluate {DomURLs\_BERT} on various malicious URLs and domain names classification tasks, including DGA, DNS tunneling, malware classification, and phishing.
    \item We conduct our experiments on both binary and multi-class classification tasks.
    \item We compare our model with several state-of-the-art deep learning models, including character-based models and pre-trained cybersecurity BERT models. 
\end{itemize}

The rest of the paper is organized as follows: Section~\ref{relatedwork} reviews related work in the field of malicious domain and URL detection. In Section~\ref{methodology}, we describe the proposed method for {DomURLs\_BERT} pre-training. Sections~\ref{results} presents the experimental results. Finally, Section~\ref{conclusion} concludes the paper and outlines potential directions for future research.

\section{Related Work}\label{relatedwork}

The field of natural language processing is currently undergoing a revolutionary transformation, driven by the advent of large pre-trained language models (PLMs) based on the groundbreaking Transformer architecture \cite{VaswaniSPUJGKP17}.  However, applying these models to domain-specific tasks poses challenges, as general models often fail to represent domain-specific terms and contexts not covered in their training data. To address this issue, domain-specific PLMs have been developed, such as BioBERT \cite{lee2020biobert} for biomedical text and SciBERT \cite{beltagy-etal-2019-scibert} for scientific literature. Similarly, in the cybersecurity domain, several models based on the BERT architecture \cite{Devlin2019BERTPO} have been created to capture domain-specific language and improve performance on cybersecurity-related tasks \cite{xu2024large}.

For instance, CyBERT \cite{ranade2021cybert} is a domain-specific variant of BERT pre-trained on a large cybersecurity corpus using MLM. It focuses on generating contextualized embeddings specifically designed for cybersecurity tasks like cyber threat intelligence and malware detection. In the same context,  CySecBERT \cite{BayerKSR24} is a domain-adapted version of BERT pre-trained on large cybersecurity corpora. It is designed to improve performance across multiple cybersecurity tasks, including classification and named entity recognition while addressing challenges like catastrophic forgetting during domain adaptation. CySecBERT has demonstrated superior performance compared to both BERT and CyBERT in several cybersecurity tasks. Additionally, SecureBERT \cite{aghaei2023securebert}, based on the RoBERTa architecture, incorporates continual pre-training with a specialized tokenizer and fine-tuned pre-trained weights to capture both general and cybersecurity-specific language. Evaluated on MLM and NER tasks, SecureBERT has shown promising results in comprehending cybersecurity text. On the other hand, SecBERT \cite{SecBERT}, developed from scratch, is trained on various cybersecurity corpora, such as "APTnotes" and "CASIE," and targets a broad range of cybersecurity data.

More recently, several models have focused on specific tasks within the cybersecurity domain. For example, MalBERT \cite{rahali2021malbert}, a BERT-based model, is specialized in detecting malicious software. Similarly, Li et al. \cite{li2024URLBERT} introduced URLBERT, the first pre-trained model specifically designed for URL classification and detection tasks. URLBERT incorporates novel pre-training techniques, such as self-supervised contrastive learning and virtual adversarial training, to enhance its understanding of URL structures and robustness, achieving state-of-the-art results in phishing detection and web page classification. Motivated by the success of PLMs in cybersecurity tasks, we propose DomURLs\_BERT, a specialized BERT-based encoder pre-trained on a large multilingual corpus of URLs, domain names, and DGA datasets.  This paper contextualizes DomURLs\_BERT by reviewing recent studies on the classification of malicious domain names and URLs. For a detailed review of existing large language models in cybersecurity, readers can refer to the recent study by Xu et al. \cite{xu2024large}.

\subsection{Malicious domain names classification}
Detecting malicious domains, especially those generated by domain generation algorithms, is a crucial task in cybersecurity. Early work by Yadav et al. \cite{yadav2010detecting} laid the foundation by focusing on detecting algorithmically generated malicious domain names through linguistic analysis. Building on this, Cucchiarelli et al. \cite{cucchiarelli2021algorithmically} proposed using n-gram features, enhancing the ability to capture linguistic patterns in DGA-generated domains. Liew and Law \cite{LiewL23} further advanced the field by introducing subword tokenization techniques for DGA classification, a method that allows more granular token analysis, improving model robustness against unseen domain variations. Shi et al. \cite{shi2018malicious} explored machine learning techniques, particularly extreme machine learning, for detecting malicious domain names. This approach demonstrates the effectiveness of using machine learning models to identify domains that exhibit abnormal patterns. Tian et al. \cite{tian2024dom} introduced Dom-bert, a pre-trained model designed to detect malicious domains, leveraging contextual information embedded in domain names to enhance detection performance. In the broader context, Kang \cite{kang2021} reviewed various malicious domain detection techniques, while Hamroun et al. \cite{hamroun2024review}
focused specifically on lexical-based methods, emphasizing the importance of features derived from the domain names themselves. Together, these works underscore the importance of both lexical features and advanced machine-learning techniques in detecting DGA-generated and malicious domains.

%\textcolor{red}{DGA related work is also here \cite{LiewL23} Use of subword tokenization for domain generation algorithm classification } \cite{yadav2010detecting} Detecting algorithmically generated malicious domain names \\ \cite{shi2018malicious} Malicious domain name detection based on extreme machine learning \\ \cite{kang2021} A Review: How to Detect Malicious Domains, 2021\\ \cite{cucchiarelli2021algorithmically} Algorithmically generated malicious domain names detection based on n-grams features\\ \cite{LiewL23} Use of subword tokenization for domain generation algorithm classification \\ \cite{hamroun2024review} A review on lexical based malicious domain name detection methods, 2024\\ \cite{tian2024dom} Dom-bert: Detecting malicious domains with pre-training model\\

\subsection{Malicious URLs classification}

In the field of malicious URLs detection, machine learning and deep learning approaches have been extensively studied and developed, with significant advancements in recent years. These approaches can be broadly categorized into traditional machine learning methods, neural network-based methods, and transformer-based models. Traditional machine learning methods, which rely on manually engineered features, were initially prominent in malicious URL detection. Sahoo et al.  \cite{sahoo2017malicious} provided a comprehensive survey of these early efforts, highlighting how machine learning techniques such as support vector machines, decision trees, naive Bayes, and random forests were applied to extract statistical and lexical features from URLs. Similarly, Aljabri et al. \cite{aljabri2022detecting} reviewed more recent methods and highlighted the shift toward deep learning techniques due to their ability to automate feature extraction and improve detection performance. 

The shift towards deep learning led to the development of several promising models. In this context, Le et al. \cite{le2018urlnet} proposed URLNet, a deep learning-based method that captures both character- and word-level representations of URLs to improve classification accuracy. Vazhayil et al.  \cite{vazhayil2018comparative} performed a comparative study between shallow and deep networks, concluding that deep networks outperform traditional machine learning models by capturing more complex patterns in URLs. Afzal et al. \cite{afzal2021urldeepdetect} took this further by introducing Urldeepdetect, a deep learning model that integrates semantic vector models to enhance URL representation.

More recently, transformer-based models have emerged as a dominant approach, driven by their capacity to understand the semantic and contextual information of URLs. As previously mentioned,  the BERT model and its variants have been particularly influential in this area. Chang et al. \cite{chang2021research} and Otieno et al.  \cite{otieno2023detecting} explored the application of BERT for URL detection, demonstrating that transformer models outperform traditional methods in terms of both accuracy and robustness. Building on these efforts, Su et al. \cite{su2023bert} and Yu et al. \cite{yu2024efficient} proposed modified BERT variants that further enhance semantic understanding for malicious URL detection. The introduction of URLBERT by Li et al. \cite{li2024URLBERT}, a contrastive and adversarial pre-trained model, continues this trend, pushing the boundaries of transformer-based URL classification. Several other transformer-based models have also been proposed, focusing on improving phishing URL detection. For example, URLTran \cite{URLTran9653028} applies transformers specifically to phishing detection, while Bozkir et al. \cite{bozkir2023grambeddings} introduced GramBeddings, a neural network that utilizes n-gram embeddings to enhance the identification of phishing URLs. This direction was further extended by Liu et al. 
 \cite{liu2023malicious}, who combined a pre-trained language model with multi-level feature attention for improved detection accuracy.

Overall, the progression from traditional machine learning approaches to advanced deep learning and transformer-based models has significantly improved the ability to classify malicious URLs. The integration of semantic understanding, n-gram embeddings, and pre-trained models has pushed the state-of-the-art, enabling more accurate and robust detection of malicious URLs across different attack types.

\section{Methodology}\label{methodology}

This section presents our methodology for pre-training the DomURLs\_BERT encoder, focusing on the collection of pre-training data, preprocessing of domain names and URLs, tokenizer training, and domain-adaptive pre-training.

\begin{table}[htbp]
  \centering
  \caption{Pre-training Dataset}
    \begin{tabular}{lcc}
\cmidrule{2-3}          & \textbf{Training} & \textbf{Development} \\
    \midrule
    \textbf{domain names} & 19,941,474 & 1,049,558 \\
    \textbf{URLs} & 355,116,387 & 18,690,330 \\
    \midrule
    \textbf{Total} & 375,057,861 & 19,739,888 \\
    \bottomrule
    \end{tabular}%
  \label{tab:pretrainindata}%
\end{table}%

\subsection{Pre-training data}
We have collected a large-scale pre-training corpus of domain names and URLs from the following datasets:
\begin{itemize}
    \item \textbf{mC4}:  The multilingual colossal Common Crawl Corpus\footnote{\url{https://hf.co/datasets/legacy-datasets/mc4}}. This is a cleaned version of the Common Crawl's web corpus, curated by the Allen Institute for Artificial Intelligence \cite{2019t5}, containing approximately 170 million URLs.
     \item \textbf{falcon-refinedweb}: An English large-scale dataset curated for large language model pre-training. This dataset is compiled from CommonCrawl, using strict filtering and extensive deduplication \cite{refinedweb}, and contains around 128 million URLs\footnote{\url{https://hf.co/datasets/tiiuae/falcon-refinedweb}}.
    \item \textbf{CBA Web tracking datasets}: A dataset compiled by the Broadband Communications Systems and Architectures Research Group\footnote{\url{https://cba.upc.edu/downloads/category/29-web-tracking-datasets\#}}, containing 76M URLs and 1.5M domain names.
    \item  \textbf{Tranco top 1M}: is a dataset of top 1M  domain names compiled and ranked by Tranco\footnote{\url{https://tranco-list.eu/}} \cite{pochat2018tranco}.
     \item \textbf{UTL\_DGA22}: A Domain Generation Algorithm botnet dataset, containing 4.3 million entries from 76 DGA families \cite{tuan2023utldga22}.
    \item \textbf{UMUDGA}: A dataset for profiling DGA-based botnets, consisting of 30 million manually labeled DGA entries \cite{ZAGO2020105400}. 
\end{itemize}

Since the pre-training dataset is curated from multiple sources, the data cleaning process includes deduplication based on exact matching. The final pre-training dataset contains 375,057,861 samples for model training and 19,739,888 samples for development. Table \ref{tab:pretrainindata} provides details on the collected dataset, which is publicly available on Hugging Face Datasets\footnote{\url{https://hf.co/datasets/amahdaouy/Web_DomURLs}}.

\subsection{Pre-training Procedure}

\subsubsection{Preprocessing}
\begin{figure}[htbp]
    \centering
    \includegraphics[width=0.75\linewidth]{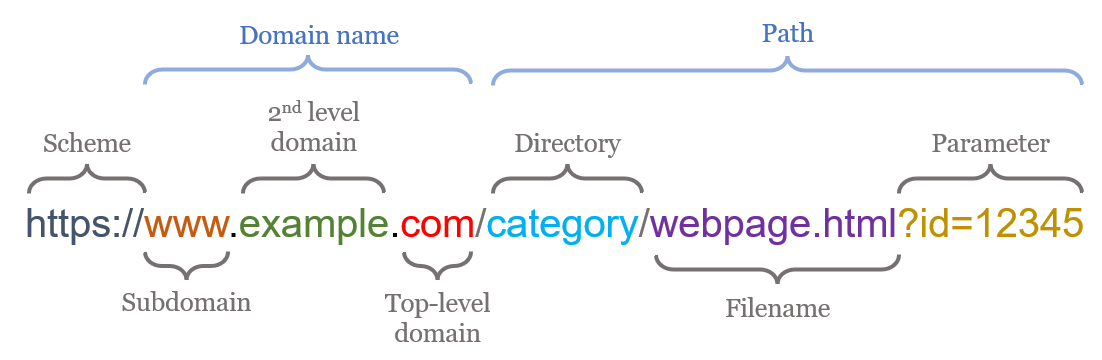}
    \caption{Overall URL structure}
    \label{fig:urlstructure}
\end{figure}

A URL consists of several components, which can be grouped into three main parts: the scheme (protocol), the domain name, and the path. Figure \ref{fig:urlstructure} illustrates the overall structure of a URL (source\footnote{\url{https://www.seoforgooglenews.com/p/everything-urls-news-publishers}}). Our proposed input preprocessing method involves removing the protocol identifier and splitting the URL into two parts: the domain name and the path. These two parts are delimited by special tokens, [DOMAIN] and [PATH], indicating the start of the domain name and the URL path, respectively. Additionally, if the input URL contains an IP address instead of a domain name, we use the [IP] and [IPv6] special tokens in place of [DOMAIN] for IPv4 and IPv6 addresses, respectively. Finally, the [CLS] and [SEP] tokens are appended to the start and end of the input URL or domain, as follows:

\begin{figure*}[!ht]
	\includegraphics[scale=0.55]{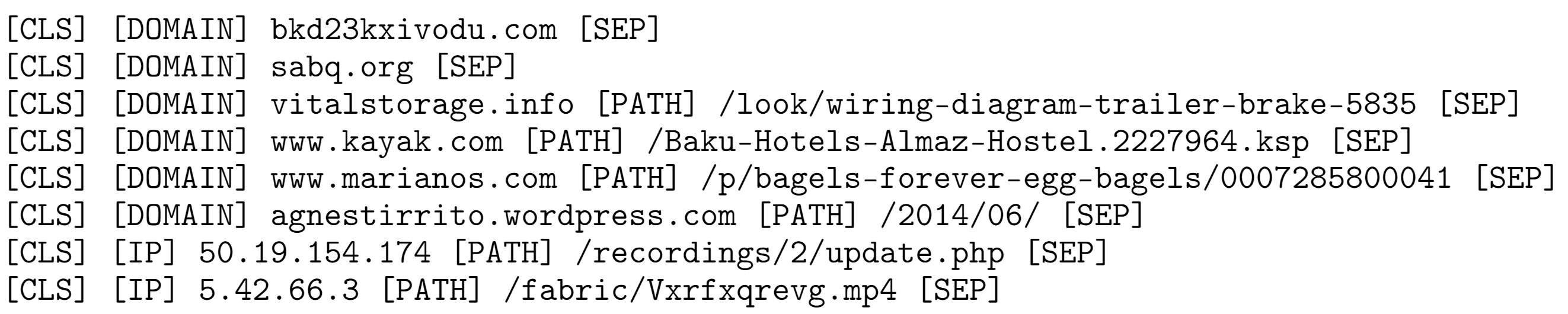}
%\begin{minted}
%[CLS] [DOMAIN] bkd23kxivodu.com [SEP]
%[CLS] [DOMAIN] sabq.org [SEP]
%[CLS] [DOMAIN] vitalstorage.info [PATH] /look/wiring-diagram-trailer-brake-5835 [SEP]
%[CLS] [DOMAIN] www.kayak.com [PATH] /Baku-Hotels-Almaz-Hostel.2227964.ksp [SEP]
%[CLS] [DOMAIN] www.marianos.com [PATH] /p/bagels-forever-egg-bagels/0007285800041 [SEP]
%[CLS] [DOMAIN] agnestirrito.wordpress.com [PATH] /2014/06/ [SEP]
%[CLS] [IP] 50.19.154.174 [PATH] /recordings/2/update.php [SEP]
%[CLS] [IP] 5.42.66.3 [PATH] /fabric/Vxrfxqrevg.mp4 [SEP]
%\end{minted}
\caption{A sample of preprocessed domain names and URLs}
\end{figure*}
\subsubsection{Tokenizer training}
After cleaning and preprocessing the data, we trained our tokenizer from scratch using the SentencePiece tokenization method, which employs the Byte Pair Encoding (BPE) algorithm \cite{kudo2018sentencepiece}. SentencePiece is language-agnostic and does not require any pre-tokenization, as it processes input as a sequence of Unicode characters. For tokenizer training, we utilized the HuggingFace tokenizers library\footnote{\url{https://github.com/huggingface/tokenizers}}. The vocabulary size was set to 32,000.

\subsubsection{Domain-adaptive pre-training}
Domain-adaptive pre-training has been shown to enhance the contextualized word embeddings of existing domain-generic Pre-trained Language Models (PLMs) \cite{gururangan2020}. This improvement has also been demonstrated in cybersecurity applications, where several domain-adapted models have been proposed \cite{aghaei2023securebert,BayerKSR24,li2024URLBERT,URLTran9653028,ICASSP2023,xu2024large}. Following this trend, we continued the pre-training of the BERT-base encoder introduced in \cite{Devlin2019BERTPO}. The model consists of approximately 110 million parameters, with 12 transformer layers, a hidden dimension size of 768, and 12 attention heads.  

Pre-training is performed using the MLM objective on our dataset, following the guidelines of Devlin et al. (2019) \cite{Devlin2019BERTPO}, where 15\% of the input sequence's subwords are randomly selected for masking. The model is trained to minimize the cross-entropy loss between the predicted sequence and the original sequence. We use the HuggingFace transformers\footnote{\url{https://github.com/huggingface/transformers}} library for training on a server equipped with 4 NVIDIA A100 GPUs, each with 80GB of RAM. The maximum sequence length, per-device batch size, and learning rate are set to 128, 768, and $1\times 10^{-4}$, respectively. The model is trained for 260,000 steps. Our pre-trained model is publicly available on HuggingFace Models\footnote{\url{https://hf.co/amahdaouy/DomURLs\_BERT}}.

\section{Experiments and Results}\label{results}

In this section, we present the evaluation datasets, the deep learning models used for comparison, the experimental settings, and the evaluation metrics. We then discuss and analyze the obtained results.

\subsection{Evaluation Datasets}

To evaluate the effectiveness of our model, we employed several domain name and URL classification datasets. For malicious domain name classification, we used the \textbf{DNS Tunneling} dataset \cite{Bubnov2019}, \textbf{UMUDGA} \cite{ZAGO2020105400}, and \textbf{UTL\_DGA22} \cite{tuan2023utldga22}. Additionally, we collected a malware domain names dataset, \textbf{ThreatFox\_MalDom}, from the ThreatFox\footnote{\url{https://threatfox.abuse.ch/}} database in June 2024. For legitimate domain names, we used the Tranco list.

For malicious URL classification, we utilized several datasets, including \textbf{Mendeley AK Singh} \cite{Singh2020}, \textbf{Kaggle Malicious URLs} \cite{siddhartha2021malicious}, \textbf{Grambedding} \cite{BOZKIR2023102964}, \textbf{LNU\_Phish} \cite{apruzzese2022mitigating}, \textbf{PhiUSIIL} \cite{PRASAD2024103545}, and \textbf{PhishCrawl} \cite{DO2024269}. We also curated a malware URL dataset, \textbf{ThreatFox\_MalURLs}, from the ThreatFox database in June 2024. For legitimate URLs, we employed the benign URLs from the \textbf{Kaggle Malicious URLs} dataset \cite{siddhartha2021malicious}.

All the used datasets have been divided into 60\%, 20\%, and 20\% for training, validation, and testing, respectively. Table \ref{tab:evaldatasets} summarizes the characteristics of the evaluation datasets.

\begin{table*}[htbp]
  \centering
  \caption{Evaluation datasets}
  \resizebox{\textwidth}{!}{% 
    \begin{tabular}{llllccccc}
    \toprule
    \textbf{Dataset} & \textbf{Type} & \textbf{Year} & \textbf{Task} & \textbf{Num Classes} & \textbf{Size} & \textbf{Training} & \textbf{Validation} & \textbf{Test} \\
    \midrule
    DNS Tunneling \cite{Bubnov2019} & domain names & 2019 & DNS Tunneling & 5     & 96,063 & 57,637 & 19,213 & 19,213 \\
    UMUDGA \cite{ZAGO2020105400} & domain names & 2020& DGA botnet & 51    & 3,098,626 & 1,859,175 & 619,725 & 619,726 \\
    UTL\_DGA22 \cite{tuan2023utldga22} & domain names & 2022 & DGA botnet & 77    & 4,297,916 & 2,578,749 & 859,583 & 859,584 \\
    ThreatFox\_MalDom (ours) & domain names & 2024 & Malware  & 65    & 176,065 & 105,639 & 35,213 & 35,213 \\
    \midrule
    Mendely AK Singh \cite{Singh2020} & URLs & 2020 & Malicious & 2     & 1,530,687 & 812,253 & 359,217 & 359,217 \\
    Kaggle malicious URLs \cite{siddhartha2021malicious} & URLs  & 2021& Malicious & 4  & 641,126 & 384,673 & 128,225 & 128,228 \\
    Grambedding \cite{BOZKIR2023102964} & URLs & 2023  & Phishing & 2     & 800,003 & 480,008 & 159,997 & 159,998 \\
    LNU\_Phish \cite{apruzzese2022mitigating} & URLs & 2022 & Phishing & 2     & 22,501 & 13,501 & 4,500  & 4,500 \\
    PhiUSIIL \cite{PRASAD2024103545} & URLs & 2024  & Phishing & 2     & 235,370 & 141,222 & 47,074 & 47,074 \\
    PhishCrawl \cite{DO2024269} & URLs  & 2024 & Phishing & 2     & 101,827 & 61,095 & 20,366 & 20,366 \\
    ThreatFox\_MalURLs (ours) & URLs & 2024 & Malware & 58    & 682,003 & 409,201 & 136,401 & 136,401 \\
    \bottomrule
    \end{tabular}%
    }%
  \label{tab:evaldatasets}%
\end{table*}%

\subsection{Comparison methods}

We compared our model with several state-of-the-art deep learning models, including six character-based RNN and CNN models, as described below:

\begin{itemize}
    \item \textbf{CharCNN}: This model employs an embedding layer followed by three one-dimensional convolutional layers with kernel sizes of 3, 4, and 5, respectively. The final convolutional layer is followed by a dropout layer and a classification layer.
    \item \textbf{CharGRU}: This model uses an embedding layer and multiple GRU layers. The last GRU layer is followed by a dropout layer and a classification layer.
    \item \textbf{CharLSTM}: This model utilizes an embedding layer and multiple LSTM layers. The final LSTM layer is followed by a dropout layer and a classification layer.
    \item \textbf{CharBiGRU}: This model uses an embedding layer and multiple bidirectional GRU layers. The last BiGRU layer is followed by a dropout layer and a classification layer.
    \item \textbf{CharBiLSTM}: This model employs an embedding layer and multiple bidirectional LSTM layers. The final BiLSTM layer is followed by a dropout layer and a classification layer.
    \item \textbf{CharCNNBiLSTM}: This model uses CNN layers to extract local features from character embeddings, which are then passed into a BiLSTM layer to capture contextual dependencies.
\end{itemize}

Moreover, we compared our model with five state-of-the-art domain-generic and domain-specific BERT-based PLMs, including \textbf{BERT} \cite{Devlin2019BERTPO}, \textbf{SecBERT}  \cite{SecBERT}, \textbf{SecureBERT} \cite{aghaei2023securebert}, \textbf{CySecBERT} \cite{BayerKSR24}, and \textbf{URLBERT} \cite{li2024URLBERT}.

\subsection{Experiments settings}

We implemented our model and the other state-of-the-art models using Pytorch\footnote{\url{https://pytorch.org/}} deep learning framework, Lightning\footnote{\url{https://lightning.ai/}}, and HuggingFace transformers\footnote{\url{https://github.com/huggingface/transformers}} library. All our experiments have been conducted on a Dell PowerEdge XE8545 server, having 4 NVIDIA A100-SXM4-80GB GPUs, 1000 GiB RAM, and 2 AMD EPYC 7713 64-Core Processor 1.9GHz.

All models are trained using AdamW optimizer \cite{LoshchilovH19}. We used a batch size of 128 and the maximum sequence length is fixed to 128 and 64 for URLs and domain names, respectively. For character-based deep learning models, the number of epochs, the learning rate, the weight decay, the number of RNN layers, the hidden dimensions size are fixed to $20$, $1 \time 10^{-3}$, $1 \time 10^{-3}$, 3, 128, respectively. For BERT-based models, the number of epochs, the learning rate, and the weight decay are fixed to $10$, $1 \time 10^{-5}$, $1 \time 10^{-3}$, respectively. For all models, weight decay is applied to all the layers weights except biases and Layer Normalization. For all models and dataset, we  utilized the following performance measures: 
\begin{itemize}
    \item \textbf{Accuracy}:  \( \text{Accuracy} = \frac{\text{TP} + \text{TN}}{\text{TP} + \text{FP} + \text{TN} + \text{FN}} \) is The proportion of all correct predictions (both true positives and true negatives) out of all predictions.

    \item \textbf{True Positive Rate (TPR) / Recall / Sensitivity/ Detection Rate}: \( \text{TPR} = \frac{\text{TP}}{\text{TP} + \text{FN}} \) is the proportion of actual positives that are correctly identified by the model.

    \item \textbf{Specificity (SPC) / True Negative Rate}: \( \text{SPC} = \frac{\text{TN}}{\text{FP} + \text{TN}} \) is the proportion of actual negatives that are correctly identified by the model.

    \item \textbf{Positive Predictive Value (PPV) / Precision}: \( \text{PPV} = \frac{\text{TP}}{\text{TP} + \text{FP}} \) is the proportion of predicted positives that are actually positive.

    \item \textbf{Negative Predictive Value (NPV)}: \( \text{NPV} = \frac{\text{TN}}{\text{TN} + \text{FN}} \) is the proportion of predicted negatives that are actually negative.

    \item \textbf{F1 score}:  \( \text{F1} = \frac{2 \cdot (\text{PPV} \times \text{TPR})} {\text{PPV} + \text{TPR}} \) is the harmonic mean of precision and recall.

    \item \textbf{Weighted F1 Score (F1\_wted)}: \( F1_{\text{weighted}} = \sum_{i=1}^{n} \left( w_i \cdot F1_i \right) \) where \( w_i \) is the proportion of the total number of samples belonging to class \( i \), and \( F1_i \) is the F1 score for class \( i \). This score considers each class's importance by weighting their respective F1 scores according to their frequency.

    \item \textbf{Micro F1 Score (F1\_mic)}: \( F1_{\text{micro}} = \frac{2 \cdot TP_{\text{micro}}}{2 \cdot TP_{\text{micro}} + FP_{\text{micro}} + FN_{\text{micro}}} \) aggregates the contributions of all classes to compute precision and recall, treating all instances equally, regardless of the class.
    
    \item \textbf{Diagnostic Efficiency (DE)}: \( \text{DE} = \text{TPR} \times \text{SPC} \) is  the product of sensitivity (TPR) and specificity (SPC). Indicates the overall diagnostic ability of the test. Higher values indicate better performance.

    \item \textbf{False Positive Rate (FPR)}: \( \text{FPR} = \frac{\text{FP}}{\text{FP} + \text{TN}} \) is the proportion of actual negatives that are incorrectly identified as positive by the model. Lower values indicate better performance.

    \item \textbf{False Discovery Rate (FDR)}: \( \text{FDR} = \frac{\text{FP}}{\text{FP} + \text{TP}} \) is the proportion of predicted positives that are actually negative. Lower values indicate better performance.

    \item \textbf{False Negative Rate (FNR)}: \( \text{FNR} = \frac{\text{FN}}{\text{FN} + \text{TP}} \) is the proportion of actual positives that are incorrectly identified as negative by the model. Lower values indicate better performance.

\end{itemize}
For all evaluation measures, we report the macro-average performances (except F1\_wted, F1\_mic, and Accuracy). 

\begin{table*}[h!]
  \centering
  \caption{Malicious domain names detection. For each dataset, the best performances are highlighted in bold font. All performance measures are presented as percentages.}
  \resizebox{\textwidth}{!}{% 
    \begin{tabular}{llcccccccccccc}
    \toprule
    \textbf{dataset} & \textbf{Name} & \textbf{Accuracy} & \textbf{F1\_macro} & \textbf{F1\_wted} & \textbf{F1\_micro} & \textbf{TPR} & \textbf{PPV} & \textbf{NPV} & \textbf{SPC} & \textbf{DE} & \textbf{FDR} & \textbf{FPR} & \textbf{FNR} \\

    \midrule
    \multirow{12}[2]{*}{\rotatebox{90}{ThreatFox\_MalDomains}} & CharBiGRU & 85.26 & 84.58 & 85.02 & 85.26 & 83.96 & 86.24 & 86.24 & 83.96 & 69.58 & 13.764 & 16.040 & 16.040 \\
          & CharBiLSTM & 85.70 & 85.24 & 85.60 & 85.70 & 84.86 & 85.94 & 85.94 & 84.86 & 71.63 & 14.061 & 15.139 & 15.139 \\
          & CharCNN & 85.03 & 84.46 & 84.87 & 85.03 & 83.99 & 85.48 & 85.48 & 83.99 & 69.97 & 14.516 & 16.006 & 16.006 \\
          & CharCNNBiLSTM & 84.94 & 84.33 & 84.75 & 84.94 & 83.82 & 85.53 & 85.53 & 83.82 & 69.58 & 14.466 & 16.181 & 16.181 \\
          & CharGRU & 85.09 & 84.61 & 84.98 & 85.09 & 84.25 & 85.26 & 85.26 & 84.25 & 70.60 & 14.736 & 15.750 & 15.750 \\
          & CharLSTM & 85.03 & 84.50 & 84.89 & 85.03 & 84.06 & 85.39 & 85.39 & 84.06 & 70.15 & 14.608 & 15.938 & 15.938 \\
          & CySecBERT & 87.04 & 86.57 & 86.91 & 87.04 & 86.08 & 87.57 & 87.57 & 86.08 & 73.60 & 12.426 & 13.917 & 13.917 \\
          & SecBERT & 84.97 & 84.63 & 84.94 & 84.97 & 84.50 & 84.79 & 84.79 & 84.50 & 71.29 & 15.213 & 15.497 & 15.497 \\
          & SecureBERT & 86.55 & 86.04 & 86.41 & 86.55 & 85.54 & 87.11 & 87.11 & 85.54 & 72.62 & 12.886 & 14.458 & 14.458 \\
          & BERT & 86.90 & 86.46 & 86.79 & 86.90 & 86.03 & 87.25 & 87.25 & 86.03 & 73.62 & 12.746 & 13.966 & 13.966 \\
          & URLBERT & 85.10 & 84.59 & 84.97 & 85.10 & 84.17 & 85.42 & 85.42 & 84.17 & 70.37 & 14.584 & 15.831 & 15.831 \\
          & DomURLs\_BERT & \textbf{88.73} & \textbf{88.33} & \textbf{88.62} & \textbf{88.73} & \textbf{87.83} & \textbf{89.30} & \textbf{89.30} & \textbf{87.83} & \textbf{76.72} & \textbf{10.705} & \textbf{12.165} & \textbf{12.165} \\
    \midrule
    \multirow{12}[2]{*}{\rotatebox{90}{UMUDGA}} & CharBiGRU & 97.52 & 97.16 & 97.52 & 97.52 & 97.23 & 97.10 & 97.10 & 97.23 & 94.52 & 2.901 & 2.775 & 2.775 \\
          & CharBiLSTM & 98.54 & 98.33 & 98.54 & 98.54 & 98.27 & 98.40 & 98.40 & 98.27 & 96.56 & 1.600 & 1.734 & 1.734 \\
          & CharCNN & 96.86 & 96.41 & 96.86 & 96.86 & 96.35 & 96.46 & 96.46 & 96.35 & 92.82 & 3.539 & 3.646 & 3.646 \\
          & CharCNNBiLSTM & 97.76 & 97.44 & 97.76 & 97.76 & 97.46 & 97.42 & 97.42 & 97.46 & 94.97 & 2.581 & 2.544 & 2.544 \\
          & CharGRU & 97.52 & 97.16 & 97.52 & 97.52 & 97.15 & 97.17 & 97.17 & 97.15 & 94.36 & 2.831 & 2.853 & 2.853 \\
          & CharLSTM & 98.35 & 98.11 & 98.34 & 98.35 & 98.05 & 98.16 & 98.16 & 98.05 & 96.13 & 1.836 & 1.951 & 1.951 \\
          & CySecBERT & 98.74 & 98.55 & 98.74 & 98.74 & 98.49 & 98.61 & 98.61 & 98.49 & 97.00 & 1.386 & 1.507 & 1.507 \\
          & SecBERT & 97.95 & 97.65 & 97.95 & 97.95 & 97.47 & 97.83 & 97.83 & 97.47 & 94.98 & 2.167 & 2.535 & 2.535 \\
          & SecureBERT & 98.81 & 98.64 & 98.81 & 98.81 & 98.59 & 98.69 & 98.69 & 98.59 & 97.19 & 1.306 & 1.414 & 1.414 \\
          & BERT & 98.76 & 98.58 & 98.76 & 98.76 & 98.59 & 98.58 & 98.58 & 98.59 & 97.20 & 1.424 & 1.407 & 1.407 \\
          & URLBERT & 97.58 & 97.24 & 97.58 & 97.58 & 97.35 & 97.13 & 97.13 & 97.35 & 94.76 & 2.866 & 2.652 & 2.652 \\
          & DomURLs\_BERT & \textbf{99.11} & \textbf{98.98} & \textbf{99.11} & \textbf{99.11} & \textbf{98.98} & \textbf{98.99} & \textbf{98.99} & \textbf{98.98} & \textbf{97.96} & \textbf{1.012} & \textbf{1.024} & \textbf{1.024} \\
    \midrule
    \multirow{12}[2]{*}{\rotatebox{90}{UTL\_DGA22}} & CharBiGRU & 97.15 & 96.03 & 97.16 & 97.15 & 96.27 & 95.79 & 95.79 & 96.27 & 92.66 & 4.206 & 3.726 & 3.726 \\
          & CharBiLSTM & 98.26 & 97.55 & 98.25 & 98.26 & 97.50 & 97.61 & 97.61 & 97.50 & 95.03 & 2.386 & 2.504 & 2.504 \\
          & CharCNN & 96.58 & 95.18 & 96.57 & 96.58 & 94.90 & 95.46 & 95.46 & 94.90 & 89.97 & 4.537 & 5.098 & 5.098 \\
          & CharCNNBiLSTM & 97.45 & 96.44 & 97.45 & 97.45 & 96.48 & 96.39 & 96.39 & 96.48 & 93.05 & 3.608 & 3.521 & 3.521 \\
          & CharGRU & 97.20 & 96.10 & 97.21 & 97.20 & 96.33 & 95.87 & 95.87 & 96.33 & 92.77 & 4.127 & 3.667 & 3.667 \\
          & CharLSTM & 98.30 & 97.61 & 98.29 & 98.30 & 97.54 & 97.68 & 97.68 & 97.54 & 95.12 & 2.320 & 2.459 & 2.459 \\
          & CySecBERT & 98.62 & 98.06 & 98.62 & 98.62 & 97.91 & 98.22 & 98.22 & 97.91 & 95.84 & 1.776 & 2.093 & 2.093 \\
          & SecBERT & 97.88 & 97.03 & 97.88 & 97.88 & 96.97 & 97.10 & 97.10 & 96.97 & 93.99 & 2.903 & 3.034 & 3.034 \\
          & SecureBERT & 98.64 & 98.09 & 98.64 & 98.64 & 98.02 & 98.17 & 98.17 & 98.02 & 96.07 & 1.834 & 1.978 & 1.978 \\
          & BERT & 98.58 & 98.00 & 98.58 & 98.58 & 97.84 & 98.17 & 98.17 & 97.84 & 95.71 & 1.831 & 2.160 & 2.160 \\
          & URLBERT & 97.55 & 96.56 & 97.55 & 97.55 & 96.55 & 96.58 & 96.58 & 96.55 & 93.18 & 3.420 & 3.451 & 3.451 \\
          & DomURLs\_BERT & \textbf{98.80} & \textbf{98.32} & \textbf{98.80} & \textbf{98.80} & \textbf{98.34} & \textbf{98.31} & \textbf{98.31} & \textbf{98.34} & \textbf{96.70} & \textbf{1.695} & \textbf{1.661} & \textbf{1.661} \\
   \midrule
    \multirow{12}[2]{*}{\rotatebox{90}{DNS Tunneling}} & CharBiGRU & 99.97 & 99.90 & 99.97 & 99.97 & 99.93 & 99.88 & 99.88 & 99.93 & 99.86 & 0.123 & 0.070 & 0.070 \\
          & CharBiLSTM & 99.96 & 99.89 & 99.96 & 99.96 & 99.93 & 99.85 & 99.85 & 99.93 & 99.85 & 0.152 & 0.073 & 0.073 \\
          & CharCNN & \textbf{99.98} & 99.95 & 99.98 & 99.98 & 99.99 & 99.91 & 99.91 & \textbf{99.99} & \textbf{99.98} & 0.088 & 0.009 & 0.009 \\
          & CharCNNBiLSTM & 99.97 & 99.92 & 99.97 & 99.97 & 99.93 & 99.91 & 99.91 & 99.93 & 99.87 & 0.093 & 0.067 & 0.067 \\
          & CharGRU & 99.97 & 99.90 & 99.97 & 99.97 & 99.90 & 99.90 & 99.90 & 99.90 & 99.81 & 0.096 & 0.096 & 0.096 \\
          & CharLSTM & 99.97 & 99.90 & 99.97 & 99.97 & 99.93 & 99.88 & 99.88 & 99.93 & 99.86 & 0.123 & 0.070 & 0.070 \\
          & CySecBERT & \textbf{99.98} & \textbf{99.94} & \textbf{99.98} & \textbf{99.98} & \textbf{99.94} & \textbf{99.94} & \textbf{99.94} & 99.94 & 99.87 & \textbf{0.064} & \textbf{0.064} & \textbf{0.064} \\
          & SecBERT & \textbf{99.98} & \textbf{99.94} & \textbf{99.98} & \textbf{99.98} & \textbf{99.94} & \textbf{99.94} & \textbf{99.94} & 99.94 & 99.87 & \textbf{0.064} & \textbf{0.064} & \textbf{0.064} \\
          & SecureBERT & \textbf{99.98} & \textbf{99.94} & \textbf{99.98} & \textbf{99.98} & \textbf{99.94} & \textbf{99.94} & \textbf{99.94} & 99.94 & 99.87 & \textbf{0.064} & \textbf{0.064} & \textbf{0.064} \\
          & BERT & \textbf{99.98} & \textbf{99.94} & \textbf{99.98} & \textbf{99.98} & \textbf{99.94} & \textbf{99.94} & \textbf{99.94} & 99.94 & 99.87 & \textbf{0.064} & \textbf{0.064} & \textbf{0.064} \\
          & URLBERT & 99.97 & 99.92 & 99.97 & 99.97 & 99.91 & 99.93 & 99.93 & 99.91 & 99.81 & 0.067 & 0.093 & 0.093 \\
          & DomURLs\_BERT & \textbf{99.98} & \textbf{99.94} & \textbf{99.98} & \textbf{99.98} & \textbf{99.94} & \textbf{99.94} & \textbf{99.94} & 99.94 & 99.87 & \textbf{0.064} & \textbf{0.064} & \textbf{0.064} \\
    \bottomrule
    \end{tabular}%
    }
  \label{tab:domcls_bin}%
\end{table*}%
\subsection{Results}

In this section, we present the results of our model alongside state-of-the-art character-based and BERT-based models. All evaluated models are compared on both binary and multi-class classification tasks for domain names and URLs. 

\subsubsection{Domains names classification Tasks}
Table \ref{tab:domcls_bin} summarizes the obtained results for binary classification of domain names. The aim of this task is to detect malicious domain names and DNS tunneling. The overall results show that DomURLs\_BERT achieves the best performance across all datasets for most evaluation metrics. However, on the DNS tunneling dataset, the CharCNN model outperforms in Specificity (SPC) and diagnostic efficiency (DE). Additionally, the results indicate that fine-tuning BERT and cybersecurity-specific BERT-based models (BERT, SecureBERT, and CySecBERT) outperforms character-based deep learning models in most datasets and metrics.

Table \ref{tab:domcls_mc} presents the obtained results for domain names multi-class classification tasks. The aim of these tasks is to classify domain names into a set of predefined class labels. The overall results show that DomURLs\_BERT model outperform the other state-of-the-art models domain generation algorithm classification datasets on most evaluation measures. Nevertheless, the CharBiLSTM achieves better F1\_macro and precision (PPV) on the UTL\_DGA22 dataset. For malware domain names classification (ThreathFox\_MalDomains dataset), DomURLs\_BERT model yields better Accuracy, F1\_wted, F1\_mic, NPV, specificity (SPC), and false positive rate (FPR). However, the best F1\_macro, precision (PPV), diagnostic efficiency (DE), recall (TPR) and  false negative rate (FNR) are achieved by  CharCNN, CharCNN, BERT, and 
CharBiGRU, respectively. For DNS tunneling, all models achieve nearly similar performances. However, the best results are obtained using CharBiGRU, CharGRU, and SecureBERT models.

\begin{table*}[htbp]
  \centering
  \caption{Malicious domain names classification. For each dataset, the best performances are highlighted in bold font. All performance measures are presented as percentages.} 
  \resizebox{\textwidth}{!}{%   

    \begin{tabular}{llcccccccccrrr}
    \toprule
    \textbf{dataset} & \textbf{Name} & \textbf{Accuracy} & \textbf{F1\_macro} & \textbf{F1\_wted} & \textbf{F1\_micro} & \textbf{TPR} & \textbf{PPV} & \textbf{NPV} & \textbf{SPC} & \textbf{DE} & \multicolumn{1}{c}{\textbf{FDR}} & \multicolumn{1}{c}{\textbf{FPR}} & \multicolumn{1}{c}{\textbf{FNR}} \\
    \midrule
    \multirow{12}[2]{*}{\rotatebox{90}{ThreatFox\_MalDomains}} & CharBiGRU & 74.87 & 35.97 & 72.80 & 74.87 & \textbf{34.31} & 46.60 & 99.50 & 99.39 & \textbf{33.85} & 34.941 & 0.611 & \textbf{65.686} \\
          & CharBiLSTM & 74.48 & 32.00 & 71.90 & 74.48 & 29.99 & 42.87 & 99.50 & 99.36 & 29.49 & 38.664 & 0.640 & 70.012 \\
          & CharCNN & 73.94 & \textbf{37.43} & 71.01 & 73.94 & 32.66 & \textbf{54.71} & 99.51 & 99.32 & 32.11 & 32.985 & 0.681 & 67.341 \\
          & CharCNNBiLSTM & 68.67 & 10.30 & 64.41 & 68.67 & 11.97 & 9.76  & 99.39 & 99.27 & 11.43 & 10.243 & 0.731 & 88.029 \\
          & CharGRU & 74.55 & 35.29 & 71.95 & 74.55 & 33.62 & 48.85 & 99.51 & 99.36 & 33.12 & 35.768 & 0.635 & 66.384 \\
          & CharLSTM & 73.53 & 29.64 & 71.21 & 73.53 & 27.99 & 37.41 & 99.47 & 99.36 & 27.52 & 30.281 & 0.636 & 72.007 \\
          & CySecBERT & 75.55 & 30.02 & 73.69 & 75.55 & 28.67 & 39.41 & 99.50 & 99.43 & 28.25 & 23.667 & 0.565 & 71.331 \\
          & SecBERT & 74.44 & 28.14 & 72.16 & 74.44 & 26.60 & 41.25 & 99.49 & 99.38 & 26.13 & 27.979 & 0.616 & 73.400 \\
          & SecureBERT & 75.88 & 31.97 & 73.85 & 75.88 & 30.88 & 39.66 & 99.52 & 99.43 & 30.44 & 23.416 & 0.573 & 69.123 \\
          & BERT & 75.36 & 25.66 & 72.83 & 75.36 & 25.21 & 35.01 & 99.52 & 99.41 & 24.74 & \textbf{17.295} & 0.587 & 74.794 \\
          & URLBERT & 72.75 & 22.91 & 69.00 & 72.75 & 21.38 & 30.98 & 99.50 & 99.29 & 20.81 & 29.021 & 0.709 & 78.624 \\
          & DomURLs\_BERT & \textbf{76.64} & 26.45 & \textbf{74.53} & \textbf{76.64} & 25.19 & 33.20 & \textbf{99.53} & \textbf{99.47} & 24.78 & 17.565 & \textbf{0.534} & 74.814 \\
    \midrule
    \multirow{12}[2]{*}{\rotatebox{90}{UMUDGA}} & CharBiGRU & 88.95 & 83.95 & 88.26 & 88.95 & 84.87 & 85.47 & 99.77 & 99.77 & 84.68 & 12.569 & 0.231 & 15.134 \\
          & CharBiLSTM & 90.68 & 86.25 & 90.23 & 90.68 & 86.62 & 87.59 & \textbf{99.81} & \textbf{99.81} & 86.47 & 10.445 & 0.192 & 13.380 \\
          & CharCNN & 85.83 & 80.68 & 85.50 & 85.83 & 80.64 & 82.67 & 99.71 & 99.70 & 80.43 & 15.367 & 0.296 & 19.365 \\
          & CharCNNBiLSTM & 87.11 & 81.38 & 86.43 & 87.11 & 82.05 & 83.51 & 99.73 & 99.73 & 81.84 & 14.528 & 0.270 & 17.947 \\
          & CharGRU & 89.26 & 84.42 & 88.54 & 89.26 & 85.25 & 86.16 & 99.78 & 99.78 & 85.07 & 11.882 & 0.224 & 14.746 \\
          & CharLSTM & 90.48 & 85.95 & 89.91 & 90.48 & 86.52 & 87.69 & 99.80 & 99.80 & 86.36 & 10.354 & 0.196 & 13.483 \\
          & CySecBERT & 90.40 & 86.07 & 90.19 & 90.40 & 86.23 & 87.36 & 99.80 & 99.80 & 86.09 & 10.681 & 0.195 & 13.767 \\
          & SecBERT & 89.48 & 85.27 & 89.29 & 89.48 & 85.59 & 86.33 & 99.78 & 99.78 & 85.44 & 11.711 & 0.216 & 14.407 \\
          & SecureBERT & 90.54 & 86.22 & 90.33 & 90.54 & 86.60 & 87.09 & \textbf{99.81} & \textbf{99.81} & 86.46 & 10.945 & 0.192 & 13.402 \\
          & BERT & 90.40 & 85.99 & 90.10 & 90.40 & 86.22 & 87.51 & 99.80 & 99.80 & 86.07 & 10.533 & 0.196 & 13.781 \\
          & URLBERT & 89.32 & 85.07 & 88.98 & 89.32 & 85.41 & 86.33 & 99.78 & 99.78 & 85.24 & 11.705 & 0.221 & 14.587 \\
          & DomURLs\_BERT & \textbf{90.86} & \textbf{86.26} & \textbf{90.46} & \textbf{90.86} & \textbf{86.66} & \textbf{87.78} & \textbf{99.81} & \textbf{99.81} & \textbf{86.52} & \textbf{10.258} & \textbf{0.185} & \textbf{13.337} \\
    \midrule
    \multirow{12}[2]{*}{\rotatebox{90}{UTL\_DGA22}} & CharBiGRU & 86.16 & 82.50 & 85.12 & 86.16 & 83.56 & 85.02 & 99.81 & 99.81 & 83.41 & 14.978 & 0.188 & 16.443 \\
          & CharBiLSTM & 88.36 & \textbf{85.23} & 87.60 & 88.36 & 86.07 & \textbf{87.05} & 99.84 & 99.84 & 85.95 & 12.952 & 0.156 & 13.933 \\
          & CharCNN & 83.13 & 79.78 & 82.69 & 83.13 & 80.05 & 82.29 & 99.77 & 99.77 & 79.90 & 17.715 & 0.228 & 19.955 \\
          & CharCNNBiLSTM & 84.46 & 80.96 & 83.84 & 84.46 & 81.43 & 83.49 & 99.79 & 99.79 & 81.28 & 16.510 & 0.210 & 18.571 \\
          & CharGRU & 86.48 & 83.42 & 85.86 & 86.48 & 84.15 & 84.90 & 99.82 & 99.82 & 84.02 & 13.805 & 0.183 & 15.850 \\
          & CharLSTM & 87.96 & 84.32 & 86.75 & 87.96 & 85.52 & 86.73 & 99.84 & 99.84 & 85.39 & 13.269 & 0.162 & 14.484 \\
          & CySecBERT & 87.97 & 84.76 & 87.35 & 87.97 & 85.35 & 86.90 & 99.84 & 99.84 & 85.23 & 13.095 & 0.160 & 14.653 \\
          & SecBERT & 86.94 & 84.16 & 86.60 & 86.94 & 84.72 & 85.76 & 99.82 & 99.83 & 84.60 & 14.235 & 0.174 & 15.281 \\
          & SecureBERT & 88.17 & 84.76 & 87.33 & 88.17 & 85.71 & 87.49 & 99.84 & 99.84 & 85.60 & \textbf{12.506} & 0.157 & 14.290 \\
          & BERT & 87.88 & 84.51 & 87.14 & 87.88 & 85.24 & 87.21 & 99.84 & 99.84 & 85.12 & 12.792 & 0.162 & 14.764 \\
          & URLBERT & 87.17 & 84.04 & 86.46 & 87.17 & 84.79 & 86.24 & 99.83 & 99.83 & 84.67 & 13.764 & 0.172 & 15.214 \\
          & DomURLs\_BERT & \textbf{88.50} & 85.09 & \textbf{87.68} & \textbf{88.50} & \textbf{86.08} & 86.08 & \textbf{99.85} & \textbf{99.85} & \textbf{85.97} & 12.624 & \textbf{0.153} & \textbf{13.918} \\
    \midrule
    \multirow{12}[2]{*}{\rotatebox{90}{DNS Tunneling}} & CharBiGRU & \textbf{99.98} & 99.94 & \textbf{99.98} & \textbf{99.98} & \textbf{99.94} & 99.93 & \textbf{100.00} & \textbf{100.00} & \textbf{99.94} & 0.067 & 0.004 & \textbf{0.056} \\
          & CharBiLSTM & 96.92 & 88.47 & 96.34 & 96.92 & 87.01 & 96.45 & 99.37 & 99.28 & 86.30 & 3.550 & 0.718 & 12.986 \\
          & CharCNN & 99.97 & 99.94 & 99.97 & 99.97 & 99.93 & 99.94 & 99.99 & 99.99 & 99.93 & 0.060 & 0.006 & 0.068 \\
          & CharCNNBiLSTM & 99.97 & 99.93 & 99.97 & 99.97 & 99.93 & 99.93 & 99.99 & 99.99 & 99.93 & 0.069 & 0.006 & 0.068 \\
          & CharGRU & \textbf{99.98} & 99.94 & \textbf{99.98} & \textbf{99.98} & \textbf{99.94} & 99.94 & \textbf{100.00} & \textbf{100.00} & 99.94 & 0.062 & 0.004 & \textbf{0.056} \\
          & CharLSTM & 96.55 & 87.65 & 96.00 & 96.55 & 85.84 & 96.09 & 99.29 & 99.19 & 85.03 & 3.912 & 0.805 & 14.163 \\
          & CySecBERT & 99.97 & 99.94 & 99.97 & 99.97 & 99.92 & 99.95 & 99.99 & 99.99 & 99.91 & 0.047 & 0.009 & 0.080 \\
          & SecBERT & 99.96 & 99.92 & 99.96 & 99.96 & 99.90 & 99.93 & 99.99 & 99.99 & 99.90 & 0.069 & 0.008 & 0.096 \\
          & SecureBERT & \textbf{99.98} & \textbf{99.95} & \textbf{99.98} & \textbf{99.98} & 99.93 & \textbf{99.96} & \textbf{100.00} & \textbf{100.00} & 99.93 & \textbf{0.042} & \textbf{0.005} & 0.065 \\
          & BERT & 99.96 & 99.91 & 99.96 & 99.96 & 99.90 & 99.92 & 99.99 & 99.99 & 99.89 & 0.081 & 0.010 & 0.096 \\
          & URLBERT & 99.93 & 99.81 & 99.93 & 99.93 & 99.84 & 99.78 & 99.98 & 99.99 & 99.83 & 0.217 & 0.015 & 0.158 \\
          & DomURLs\_BERT & 99.97 & 99.93 & 99.97 & 99.97 & 99.93 & 99.94 & 99.99 & 99.99 & 99.92 & 0.063 & 0.006 & 0.072 \\
    \bottomrule
    \end{tabular}%
      \label{tab:domcls_mc}%
    }%
\end{table*}%

\subsubsection{URLs classification Tasks}

Tables \ref{tab:urlcls_bin_p1} and \ref{tab:urlcls_bin_p2} summarizes the obtained results for URLs binary classification tasks (malicious URLs detection). The overall obtained results show that DomURLs\_BERT outperforms the state-of-the-art character-based models and 
 the evaluated BERT-based models on Grambedding, PhishCrawl, and kaggle malicious urls datasets. Besides, it achieves comparable or nearly similar performances on LNU\_Phish, Mendely AK Singh, and PhiUSIIL datasets. For LNU\_Phish, the top performances are achieved by CharGRU, CySecBERT, SecBERT, SecureBERT, BERT, and DomUrlsBERT models.  For Mendely AK Singh dataset, DomURLs\_BERT yields the best accuracy, F1\_macro, F1\_wted, and F1\_micro. Whereas, SecureBERT obtains better recall (TPR), specificity (SPC), diagnostic efficiency (DE), false positive rate (FPR) and false negative rate (FNR) performances. Additionally, the CharCNN model yields the best precision (PPV), NPV, and false detection rate (FDR). For PhiUSIIL dataset, CySecBERT achieves the best overall performances, while the other models,including DomURLs\_BERT, obtain comparable results. For ThreatFox\_MalURLs dataset, all models yield nearly perfect performances, while the top results are achieved by CySecBERT, SecureBERT, and BERT models. In accordance with domain names classification tasks, BERT model and the other cybersecurity BERT-based models obtain state-of-the-art perfornaces on all datasets.

\begin{table*}[htbp]
  \centering
  \caption{Malicious URLs detection (part 1). For each dataset, the best performances are highlighted in bold font. All performance measures are presented as percentages.}
    \resizebox{\textwidth}{!}{% 
    \begin{tabular}{llcccccccccccc}
    \toprule
    \textbf{dataset} & \textbf{Name} & \textbf{Accuracy} & \textbf{F1\_macro} & \textbf{F1\_wted} & \textbf{F1\_micro} & \textbf{TPR} & \textbf{PPV} & \textbf{NPV} & \textbf{SPC} & \textbf{DE} & \textbf{FDR} & \textbf{FPR} & \textbf{FNR} \\
    \midrule
    \multirow{12}[2]{*}{\rotatebox{90}{Grambedding}} & CharBiGRU & 97.50 & 97.50 & 97.50 & 97.50 & 97.50 & 97.51 & 97.51 & 97.50 & 95.06 & 2.489 & 2.497 & 2.497 \\
          & CharBiLSTM & 97.53 & 97.53 & 97.53 & 97.53 & 97.53 & 97.54 & 97.54 & 97.53 & 95.11 & 2.455 & 2.472 & 2.472 \\
          & CharCNN & 97.02 & 97.02 & 97.02 & 97.02 & 97.02 & 97.02 & 97.02 & 97.02 & 94.13 & 2.975 & 2.977 & 2.977 \\
          & CharCNNBiLSTM & 96.95 & 96.95 & 96.95 & 96.95 & 96.95 & 96.96 & 96.96 & 96.95 & 93.99 & 3.038 & 3.049 & 3.049 \\
          & CharGRU & 97.52 & 97.52 & 97.52 & 97.52 & 97.52 & 97.53 & 97.53 & 97.52 & 95.10 & 2.475 & 2.480 & 2.480 \\
          & CharLSTM & 97.45 & 97.45 & 97.45 & 97.45 & 97.45 & 97.46 & 97.46 & 97.45 & 94.97 & 2.542 & 2.546 & 2.546 \\
          & CySecBERT & 98.40 & 98.40 & 98.40 & 98.40 & 98.40 & 98.41 & 98.41 & 98.40 & 96.81 & 1.593 & 1.603 & 1.603 \\
          & SecBERT & 97.91 & 97.91 & 97.91 & 97.91 & 97.91 & 97.91 & 97.91 & 97.91 & 95.85 & 2.087 & 2.094 & 2.094 \\
          & SecureBERT & 98.42 & 98.42 & 98.42 & 98.42 & 98.42 & 98.42 & 98.42 & 98.42 & 96.85 & 1.581 & 1.584 & 1.584 \\
          & BERT & 98.26 & 98.26 & 98.26 & 98.26 & 98.26 & 98.26 & 98.26 & 98.26 & 96.55 & 1.737 & 1.739 & 1.739 \\
          & URLBERT & 96.87 & 96.87 & 96.87 & 96.87 & 96.87 & 96.88 & 96.88 & 96.87 & 93.84 & 3.120 & 3.128 & 3.128 \\
          & DomURLs\_BERT & \textbf{98.51} & \textbf{98.51} & \textbf{98.51} & \textbf{98.51} & \textbf{98.51} & \textbf{98.51} & \textbf{98.51} & \textbf{98.51} & \textbf{97.05} & \textbf{1.488} & \textbf{1.488} & \textbf{1.488} \\
    \midrule
    \multirow{12}[2]{*}{\rotatebox{90}{LNU\_Phish}} & CharBiGRU & 99.98 & 99.97 & 99.98 & 99.98 & 99.98 & 99.96 & 99.96 & 99.98 & 99.97 & 0.036 & 0.016 & 0.016 \\
          & CharBiLSTM & 99.98 & 99.97 & 99.98 & 99.98 & 99.98 & 99.96 & 99.96 & 99.98 & 99.97 & 0.036 & 0.016 & 0.016 \\
          & CharCNN & 99.98 & 99.97 & 99.98 & 99.98 & 99.98 & 99.96 & 99.96 & 99.98 & 99.97 & 0.036 & 0.016 & 0.016 \\
          & CharCNNBiLSTM & 99.98 & 99.97 & 99.98 & 99.98 & 99.98 & 99.96 & 99.96 & 99.98 & 99.97 & 0.036 & 0.016 & 0.016 \\
          & CharGRU & \textbf{100.00} & \textbf{100.00} & \textbf{100.00} & \textbf{100.00} & \textbf{100.00} & \textbf{100.00} & \textbf{100.00} & \textbf{100.00} & \textbf{100.00} & \textbf{0.000} & \textbf{0.000} & \textbf{0.000} \\
          & CharLSTM & 99.98 & 99.97 & 99.98 & 99.98 & 99.98 & 99.96 & 99.96 & 99.98 & 99.97 & 0.036 & 0.016 & 0.016 \\
          & CySecBERT & \textbf{100.00} & \textbf{100.00} & \textbf{100.00} & \textbf{100.00} & \textbf{100.00} & \textbf{100.00} & \textbf{100.00} & \textbf{100.00} & \textbf{100.00} & \textbf{0.000} & \textbf{0.000} & \textbf{0.000} \\
          & SecBERT & \textbf{100.00} & \textbf{100.00} & \textbf{100.00} & \textbf{100.00} & \textbf{100.00} & \textbf{100.00} & \textbf{100.00} & \textbf{100.00} & \textbf{100.00} & \textbf{0.000} & \textbf{0.000} & \textbf{0.000} \\
          & SecureBERT & \textbf{100.00} & \textbf{100.00} & \textbf{100.00} & \textbf{100.00} & \textbf{100.00} & \textbf{100.00} & \textbf{100.00} & \textbf{100.00} & \textbf{100.00} & \textbf{0.000} & \textbf{0.000} & \textbf{0.000} \\
          & BERT & \textbf{100.00} & \textbf{100.00} & \textbf{100.00} & \textbf{100.00} & \textbf{100.00} & \textbf{100.00} & \textbf{100.00} & \textbf{100.00} & \textbf{100.00} & \textbf{0.000} & \textbf{0.000} & \textbf{0.000} \\
          & URLBERT & 99.98 & 99.97 & 99.98 & 99.98 & 99.98 & 99.96 & 99.96 & 99.98 & 99.97 & 0.036 & 0.016 & 0.016 \\
          & DomURLs\_BERT & \textbf{100.00} & \textbf{100.00} & \textbf{100.00} & \textbf{100.00} & \textbf{100.00} & \textbf{100.00} & \textbf{100.00} & \textbf{100.00} & \textbf{100.00} & \textbf{0.000} & \textbf{0.000} & \textbf{0.000} \\
    \midrule
    \multirow{12}[2]{*}{\rotatebox{90}{Mendeley AK Singh}} & CharBiGRU & 98.82 & 83.53 & 98.69 & 98.82 & 77.42 & 93.41 & 93.41 & 77.42 & 54.91 & 6.589 & 22.584 & 22.584 \\
          & CharBiLSTM & 98.88 & 83.99 & 98.74 & 98.88 & 77.18 & 95.71 & 95.71 & 77.18 & 54.40 & 4.286 & 22.821 & 22.821 \\
          & CharCNN & 98.80 & 82.20 & 98.62 & 98.80 & 74.96 & \textbf{95.88} & \textbf{95.88} & 74.96 & 49.96 & \textbf{4.124} & 25.041 & 25.041 \\
          & CharCNNBiLSTM & 98.76 & 81.46 & 98.57 & 98.76 & 74.15 & 95.69 & 95.69 & 74.15 & 48.35 & 4.308 & 25.850 & 25.850 \\
          & CharGRU & 98.85 & 83.49 & 98.70 & 98.85 & 76.74 & 95.15 & 95.15 & 76.74 & 53.54 & 4.851 & 23.257 & 23.257 \\
          & CharLSTM & 98.83 & 83.12 & 98.68 & 98.83 & 76.30 & 95.14 & 95.14 & 76.30 & 52.65 & 4.859 & 23.704 & 23.704 \\
          & CySecBERT & 98.97 & 86.40 & 98.89 & 98.97 & 81.39 & 93.44 & 93.44 & 81.39 & 62.85 & 6.563 & 18.611 & 18.611 \\
          & SecBERT & 98.81 & 83.83 & 98.69 & 98.81 & 78.60 & 91.59 & 91.59 & 78.60 & 57.30 & 8.410 & 21.401 & 21.401 \\
          & SecureBERT & 98.89 & 85.64 & 98.82 & 98.89 & \textbf{81.58} & 90.98 & 90.98 & \textbf{81.58} & \textbf{63.27} & 9.017 & \textbf{18.418} & \textbf{18.418} \\
          & BERT & 98.97 & 86.30 & 98.88 & 98.97 & 81.33 & 93.27 & 93.27 & 81.33 & 62.74 & 6.725 & 18.670 & 18.670 \\
          & URLBERT & 98.68 & 80.38 & 98.48 & 98.68 & 73.49 & 93.51 & 93.51 & 73.49 & 47.06 & 6.489 & 26.506 & 26.506 \\
          & DomURLs\_BERT & \textbf{99.00} & \textbf{86.70} & \textbf{98.92} & \textbf{99.00} & 81.46 & 94.16 & 94.16 & 81.46 & 62.99 & 5.841 & 18.535 & 18.535 \\
    \midrule
    \multirow{12}[2]{*}{\rotatebox{90}{PhiUSIIL}} & CharBiGRU & 99.80 & 99.79 & 99.80 & 99.80 & 99.76 & 99.82 & 99.82 & 99.76 & 99.53 & 0.179 & 0.237 & 0.237 \\
          & CharBiLSTM & 99.78 & 99.78 & 99.78 & 99.78 & 99.75 & 99.80 & 99.80 & 99.75 & 99.51 & 0.196 & 0.247 & 0.247 \\
          & CharCNN & 99.79 & 99.79 & 99.79 & 99.79 & 99.76 & 99.82 & 99.82 & 99.76 & 99.52 & 0.182 & 0.239 & 0.239 \\
          & CharCNNBiLSTM & 99.81 & \textbf{99.81} & 99.81 & 99.81 & \textbf{99.78} & 99.83 & 99.83 & \textbf{99.78} & \textbf{99.56} & 0.167 & 0.219 & 0.219 \\
          & CharGRU & 99.79 & 99.78 & 99.79 & 99.79 & 99.76 & 99.82 & 99.82 & 99.76 & 99.51 & 0.185 & 0.244 & 0.244 \\
          & CharLSTM & 99.78 & 99.78 & 99.78 & 99.78 & 99.75 & 99.81 & 99.81 & 99.75 & 99.50 & 0.192 & 0.251 & 0.251 \\
          & CySecBERT & \textbf{99.82} & \textbf{99.81} & \textbf{99.82} & \textbf{99.82} & \textbf{99.78} & \textbf{99.84} & \textbf{99.84} & \textbf{99.78} & 99.57 & \textbf{0.161} & \textbf{0.216} & \textbf{0.216} \\
          & SecBERT & 99.80 & 99.79 & 99.80 & 99.80 & 99.76 & 99.82 & 99.82 & 99.76 & 99.53 & 0.180 & 0.236 & 0.236 \\
          & SecureBERT & 99.81 & 99.80 & 99.81 & 99.81 & \textbf{99.78} & 99.83 & 99.83 & \textbf{99.78} & \textbf{99.56} & 0.173 & 0.222 & 0.222 \\
          & BERT & 99.80 & 99.80 & 99.80 & 99.80 & \textbf{99.78} & 99.82 & 99.82 & \textbf{99.78} & 99.55 & 0.180 & 0.223 & 0.223 \\
          & URLBERT & 99.79 & 99.78 & 99.79 & 99.79 & 99.76 & 99.81 & 99.81 & 99.76 & 99.51 & 0.191 & 0.243 & 0.243 \\
          & DomURLs\_BERT & 99.80 & 99.80 & 99.80 & 99.80 & 99.77 & 99.83 & 99.83 & 99.77 & 99.54 & 0.170 & 0.229 & 0.229 \\
    \bottomrule
    \end{tabular}%
    }
  \label{tab:urlcls_bin_p1}%
\end{table*}%

\begin{table*}[htbp]
  \centering
  \caption{Malicious URLs detection (part 2). For each dataset, the best performances are highlighted in bold font. All performance measures are presented as percentages.}
    \resizebox{\textwidth}{!}{% 
    \begin{tabular}{llcccccccccccc}
    \toprule
    \textbf{dataset} & \textbf{Name} & \textbf{Accuracy} & \textbf{F1\_macro} & \textbf{F1\_wted} & \textbf{F1\_micro} & \textbf{TPR} & \textbf{PPV} & \textbf{NPV} & \textbf{SPC} & \textbf{DE} & \textbf{FDR} & \textbf{FPR} & \textbf{FNR} \\
    \midrule
    \multirow{12}[2]{*}{\rotatebox{90}{PhishCrawl}} & CharBiGRU & 96.80 & 96.78 & 96.80 & 96.80 & 96.81 & 96.75 & 96.75 & 96.81 & 93.71 & 3.251 & 3.195 & 3.195 \\
          & CharBiLSTM & 96.88 & 96.84 & 96.87 & 96.88 & 96.71 & 97.04 & 97.04 & 96.71 & 93.48 & 2.964 & 3.295 & 3.295 \\
          & CharCNN & 97.13 & 97.10 & 97.13 & 97.13 & 97.04 & 97.17 & 97.17 & 97.04 & 94.16 & 2.828 & 2.959 & 2.959 \\
          & CharCNNBiLSTM & 96.97 & 96.94 & 96.97 & 96.97 & 96.83 & 97.08 & 97.08 & 96.83 & 93.74 & 2.921 & 3.167 & 3.167 \\
          & CharGRU & 96.94 & 96.90 & 96.93 & 96.94 & 96.78 & 97.07 & 97.07 & 96.78 & 93.64 & 2.932 & 3.218 & 3.218 \\
          & CharLSTM & 97.09 & 97.06 & 97.08 & 97.09 & 96.96 & 97.19 & 97.19 & 96.96 & 93.98 & 2.807 & 3.044 & 3.044 \\
          & CySecBERT & 98.15 & 98.14 & 98.15 & 98.15 & 98.06 & 98.22 & 98.22 & 98.06 & 96.15 & 1.775 & 1.937 & 1.937 \\
          & SecBERT & 97.36 & 97.34 & 97.36 & 97.36 & 97.27 & 97.42 & 97.42 & 97.27 & 94.61 & 2.583 & 2.727 & 2.727 \\
          & SecureBERT & 98.23 & 98.22 & 98.23 & 98.23 & 98.16 & 98.28 & 98.28 & 98.16 & 96.35 & 1.722 & 1.837 & 1.837 \\
          & BERT & 98.14 & 98.13 & 98.14 & 98.14 & 98.06 & 98.21 & 98.21 & 98.06 & 96.14 & 1.792 & 1.942 & 1.942 \\
          & URLBERT & 96.70 & 96.66 & 96.69 & 96.70 & 96.57 & 96.78 & 96.78 & 96.57 & 93.25 & 3.224 & 3.426 & 3.426 \\
          & DomURLs\_BERT & \textbf{98.38} & \textbf{98.37} & \textbf{98.38} & \textbf{98.38} & \textbf{98.38} & \textbf{98.36} & \textbf{98.36} & \textbf{98.38} & \textbf{96.79} & \textbf{1.636} & \textbf{1.619} & \textbf{1.619} \\
    \midrule
    \multirow{12}[2]{*}{\rotatebox{90}{ThreatFox\_MalURLs}} & CharBiGRU & 99.98 & 99.98 & 99.98 & 99.98 & 99.98 & 99.98 & 99.98 & 99.98 & 99.96 & 0.017 & 0.019 & 0.019 \\
          & CharBiLSTM & 99.95 & 99.95 & 99.95 & 99.95 & 99.95 & 99.96 & 99.96 & 99.95 & 99.89 & 0.044 & 0.053 & 0.053 \\
          & CharCNN & 99.97 & 99.97 & 99.97 & 99.97 & 99.97 & 99.96 & 99.96 & 99.97 & 99.94 & 0.038 & 0.030 & 0.030 \\
          & CharCNNBiLSTM & 99.97 & 99.97 & 99.97 & 99.97 & 99.97 & 99.97 & 99.97 & 99.97 & 99.93 & 0.033 & 0.033 & 0.033 \\
          & CharGRU & 99.97 & 99.96 & 99.97 & 99.97 & 99.96 & 99.97 & 99.97 & 99.96 & 99.92 & 0.030 & 0.040 & 0.040 \\
          & CharLSTM & 99.96 & 99.96 & 99.96 & 99.96 & 99.96 & 99.96 & 99.96 & 99.96 & 99.91 & 0.037 & 0.045 & 0.045 \\
          & CySecBERT & \textbf{100.00} & \textbf{100.00} & \textbf{100.00} & \textbf{100.00} & \textbf{100.00} & \textbf{100.00} & \textbf{100.00} & \textbf{100.00} & 99.99 & 0.004 & 0.004 & 0.004 \\
          & SecBERT & 99.99 & 99.99 & 99.99 & 99.99 & 100.00 & 99.99 & 99.99 & 100.00 & 99.99 & 0.006 & 0.005 & 0.005 \\
          & SecureBERT & \textbf{100.00} & \textbf{100.00} & \textbf{100.00} & \textbf{100.00} & 99.99 & \textbf{100.00} & \textbf{100.00} & 99.99 & 99.99 & 0.004 & 0.005 & 0.005 \\
          & BERT & \textbf{100.00} & \textbf{100.00} & \textbf{100.00} & \textbf{100.00} & \textbf{100.00} & \textbf{100.00} & \textbf{100.00} & \textbf{100.00} & \textbf{100.00} & \textbf{0.001} & \textbf{0.002} & \textbf{0.002} \\
          & URLBERT & 99.99 & 99.99 & 99.99 & 99.99 & 99.99 & 99.99 & 99.99 & 99.99 & 99.98 & 0.009 & 0.010 & 0.010 \\
          & DomURLs\_BERT & 99.99 & 99.99 & 99.99 & 99.99 & 99.99 & 99.99 & 99.99 & 99.99 & 99.99 & 0.007 & 0.007 & 0.007 \\
    \midrule
    \multirow{12}[2]{*}{\rotatebox{90}{Kaggle malicious URLs}} & CharBiGRU & 98.68 & 98.51 & 98.68 & 98.68 & 98.28 & 98.74 & 98.74 & 98.28 & 96.58 & 1.263 & 1.716 & 1.716 \\
          & CharBiLSTM & 98.63 & 98.45 & 98.63 & 98.63 & 98.26 & 98.65 & 98.65 & 98.26 & 96.54 & 1.354 & 1.738 & 1.738 \\
          & CharCNN & 98.39 & 98.18 & 98.39 & 98.39 & 98.05 & 98.31 & 98.31 & 98.05 & 96.14 & 1.687 & 1.945 & 1.945 \\
          & CharCNNBiLSTM & 98.42 & 98.20 & 98.41 & 98.42 & 97.90 & 98.53 & 98.53 & 97.90 & 95.81 & 1.468 & 2.105 & 2.105 \\
          & CharGRU & 98.65 & 98.48 & 98.65 & 98.65 & 98.38 & 98.58 & 98.58 & 98.38 & 96.77 & 1.415 & 1.624 & 1.624 \\
          & CharLSTM & 98.58 & 98.40 & 98.58 & 98.58 & 98.29 & 98.51 & 98.51 & 98.29 & 96.60 & 1.493 & 1.712 & 1.712 \\
          & CySecBERT & 99.03 & 98.91 & 99.03 & 99.03 & 98.89 & 98.93 & 98.93 & 98.89 & 97.78 & 1.067 & 1.113 & 1.113 \\
          & SecBERT & 98.84 & 98.69 & 98.84 & 98.84 & 98.72 & 98.67 & 98.67 & 98.72 & 97.45 & 1.329 & 1.283 & 1.283 \\
          & SecureBERT & 99.00 & 98.88 & 99.00 & 99.00 & 98.86 & 98.90 & 98.90 & 98.86 & 97.73 & 1.103 & 1.140 & 1.140 \\
          & BERT & 98.95 & 98.82 & 98.95 & 98.95 & 98.90 & 98.74 & 98.74 & 98.90 & 97.81 & 1.261 & 1.101 & 1.101 \\
          & URLBERT & 98.60 & 98.41 & 98.59 & 98.60 & 98.21 & 98.62 & 98.62 & 98.21 & 96.44 & 1.379 & 1.789 & 1.789 \\
          & DomURLs\_BERT & \textbf{99.13} & \textbf{99.02} & \textbf{99.13} & \textbf{99.13} & \textbf{98.93} & \textbf{99.11} & \textbf{99.11} & \textbf{98.93} & \textbf{97.88} & \textbf{0.889} & \textbf{1.066} & \textbf{1.066} \\
    \bottomrule
    \end{tabular}%
    }
  \label{tab:urlcls_bin_p2}%
\end{table*}%

\begin{table}[htbp]
  \centering
  \caption{Malicious URLs classification. For each dataset, the best performances are highlighted in bold font.  All performance measures are presented as percentages.}
      \resizebox{\textwidth}{!}{% 

    \begin{tabular}{llrrrrrrrrrrrr}
    \toprule
    \textbf{dataset} & \textbf{Name} & \multicolumn{1}{c}{\textbf{Accuracy}} & \multicolumn{1}{c}{\textbf{F1\_macro}} & \multicolumn{1}{c}{\textbf{F1\_wted}} & \multicolumn{1}{c}{\textbf{F1\_micro}} & \multicolumn{1}{c}{\textbf{TPR}} & \multicolumn{1}{c}{\textbf{PPV}} & \multicolumn{1}{c}{\textbf{NPV}} & \multicolumn{1}{c}{\textbf{SPC}} & \multicolumn{1}{c}{\textbf{DE}} & \multicolumn{1}{c}{\textbf{FDR}} & \multicolumn{1}{c}{\textbf{FPR}} & \multicolumn{1}{c}{\textbf{FNR}} \\

    \midrule
    \multirow{12}[2]{*}{\rotatebox{90}{Kaggle malicious URLs}} & CharBiGRU & 98.36 & 97.45 & 98.35 & 98.36 & 96.86 & 98.06 & 99.29 & 99.17 & 96.05 & 1.938 & 0.832 & 3.136 \\
          & CharBiLSTM & 98.35 & 97.38 & 98.34 & 98.35 & 96.87 & 97.93 & 99.28 & 99.16 & 96.04 & 2.074 & 0.841 & 3.135 \\
          & CharCNN & 97.90 & 96.85 & 97.90 & 97.90 & 96.48 & 97.25 & 99.03 & 98.95 & 95.45 & 2.754 & 1.051 & 3.525 \\
          & CharCNNBiLSTM & 98.13 & 97.01 & 98.13 & 98.13 & 96.76 & 97.27 & 99.15 & 99.07 & 95.85 & 2.729 & 0.929 & 3.242 \\
          & CharGRU & 98.20 & 97.28 & 98.19 & 98.20 & 96.52 & 98.07 & 99.26 & 99.03 & 95.57 & 1.926 & 0.972 & 3.483 \\
          & CharLSTM & 98.39 & 97.39 & 98.38 & 98.39 & 96.69 & 98.13 & 99.36 & 99.12 & 95.83 & 1.872 & 0.876 & 3.310 \\
          & CySecBERT & 98.66 & 97.90 & 98.66 & 98.66 & 97.63 & 98.19 & 99.36 & 99.37 & 97.01 & 1.809 & 0.628 & 2.371 \\
          & SecBERT & 98.44 & 97.58 & 98.43 & 98.44 & 97.03 & 98.15 & 99.32 & 99.20 & 96.25 & 1.849 & 0.801 & 2.969 \\
          & SecureBERT & 98.81 & \textbf{98.14} & 98.81 & 98.81 & \textbf{97.76} & \textbf{98.52} & 99.47 & 99.40 & \textbf{97.17} & \textbf{1.477} & 0.598 & \textbf{2.238} \\
          & BERT & 98.65 & 97.84 & 98.65 & 98.65 & 97.64 & 98.04 & 99.35 & 99.39 & 97.04 & 1.963 & 0.612 & 2.356 \\
          & URLBERT & 98.30 & 97.32 & 98.30 & 98.30 & 96.87 & 97.78 & 99.23 & 99.16 & 96.04 & 2.216 & 0.841 & 3.133 \\
          & DomURLs\_BERT & \textbf{98.82} & 98.04 & \textbf{98.82} & \textbf{98.82} & 97.64 & 98.46 & \textbf{99.50} & \textbf{99.42} & 97.07 & 1.543 & \textbf{0.578} & 2.359 \\
    \midrule
    \multirow{12}[2]{*}{\rotatebox{90}{ThreatFox\_MalURLs}} & CharBiGRU & 98.86 & 80.15 & 98.81 & 98.86 & 77.81 & 84.97 & 99.98 & 99.98 & 77.80 & 13.310 & 0.021 & 22.186 \\
          & CharBiLSTM & 98.89 & 79.43 & 98.86 & 98.89 & 77.40 & 84.48 & 99.98 & 99.98 & 77.38 & 12.075 & 0.020 & 22.600 \\
          & CharCNN & 98.64 & 79.46 & 98.60 & 98.64 & 76.75 & 84.24 & 99.98 & 99.97 & 76.73 & 15.763 & 0.026 & 23.252 \\
          & CharCNNBiLSTM & 98.24 & 67.66 & 98.09 & 98.24 & 64.90 & 75.86 & 99.97 & 99.97 & 64.87 & 20.696 & 0.033 & 35.098 \\
          & CharGRU & 98.77 & 79.24 & 98.74 & 98.77 & 78.14 & 82.57 & 99.98 & 99.98 & 78.12 & 17.429 & 0.022 & 21.858 \\
          & CharLSTM & 98.69 & 76.37 & 98.64 & 98.69 & 75.09 & 81.70 & 99.98 & 99.98 & 75.07 & 16.580 & 0.024 & 24.907 \\
          & CySecBERT & \textbf{99.17} & \textbf{84.40} & \textbf{99.15} & \textbf{99.17} & \textbf{83.56} & \textbf{87.80} & \textbf{99.99} & \textbf{99.98} & \textbf{83.55} & 12.201 & \textbf{0.015} & \textbf{16.437} \\
          & SecBERT & 99.00 & 81.40 & 98.95 & 99.00 & 79.58 & 86.18 & 99.98 & 99.98 & 79.57 & 12.093 & 0.018 & 20.418 \\
          & SecureBERT & 99.11 & 83.54 & 99.10 & 99.11 & 81.62 & 87.68 & 99.98 & 99.98 & 81.61 & \textbf{10.599} & 0.016 & 18.380 \\
          & BERT & 99.13 & 83.06 & 99.11 & 99.13 & 82.08 & 86.05 & 99.98 & 99.98 & 82.06 & 13.954 & 0.016 & 17.924 \\
          & URLBERT & 98.51 & 75.07 & 98.47 & 98.51 & 74.00 & 79.99 & 99.97 & 99.97 & 73.98 & 18.289 & 0.027 & 25.996 \\
          & DomURLs\_BERT & 99.08 & 81.32 & 99.04 & 99.08 & 79.95 & 84.78 & 99.98 & 99.98 & 79.93 & 13.500 & 0.017 & 20.053 \\
    \bottomrule
    \end{tabular}%
    }
  \label{tab:urlcls_mc}%
\end{table}%

Table \ref{tab:urlcls_mc} summarizes the obtained results for malicious URLs multi-class classification. The overall obtained results show that DomURLs\_BERT achieve comparable results to the best performing models (SecureBERT and CySecBERT). For kaggle malicious urls dataset, DomURLs\_BERT yields slightly better Accuracy, F1\_wted, F1\_micro, NPV, specificity (SPC), and false positive rate (FPR) performances, while SecureBERT obtains slightly better F1\_macro, recall (TPR), precison, diagnostic efficiency (DE), false detection rate (FDR), and false negative rate (FNR). Although CySecBERT outperforms all evaluated models on the ThreatFox\_MalURLs dataset, DomURLs\_BERT, BERT, SecureBERT also demonstrate performances that are closely comparable. In accordance with the previously reported results, most BERT-based models, especially those adapted to cybersecurity domain, yields state-of-the-art performances on malicious URLs classification tasks.  

\section{Conclusion}\label{conclusion}

In this work, we introduced DomURLs\_BERT, a novel state-of-the-art pre-trained language model for detecting and classifying malicious or suspicious domain names and URLs. DomURLs\_BERT is pre-trained using the masked language modeling objective on a large-scale, multilingual corpus comprising URLs, domain names, and domain generation algorithm datasets. We presented a detailed methodology for data collection, preprocessing, and domain-adaptive pre-training.

To evaluate the performance of our model, we conducted experiments on several binary and multi-class classification tasks related to domain names and URLs, including phishing and malware detection, DGA identification, and DNS tunneling. Our results demonstrate that DomURLs\_BERT achieves state-of-the-art performance across multiple datasets. Furthermore, the findings highlight that character-based deep learning models, such as RNNs, CNNs, and their combinations, serve as strong end-to-end baselines for URL and domain name classification. In comparison, fine-tuning a domain-generic pre-trained BERT model and adapting BERT-based models to the cybersecurity domain consistently outperforms the baseline character-based models on most benchmark datasets. Future work includes evaluating the model performance on other domain names and URLs classification tasks and exploring robust fine-tuning approaches for dealing with adversarial attacks.

\bibliographystyle{unsrt}  

%%% Comment out this section when you 
\bibliography{main}

\end{document}